# A System for Melodic Harmonization using Schoenberg Regions, Giant Steps and Church Modes

Frederick (Tony) Fernandes

Independent Research Project

June 8th, 2022 to August 18th, 2022

Github: Link

YouTube: Link

Email: fdesanton at gmail dot com

## Abstract

Systems such as Microsoft Songsmith automatically assign chords (harmony) to a melody by minimizing the dissonance across all chord changes. Although this produces harmonious music, it is not what practicing musicians do. In this paper, I describe Harmonizer, a prototype system for melodic harmonization. Harmonizer uses Schoenberg's chart of regions as the underlying data structure that allows harmonization using several different methods. Because the chart reveals inter-chordal relationships, the harmonizations may be programmed to emphasize desired relationships. In the prototype Harmonizer, I also explore recent signal-processing methods that enable songwriters to easily input a melody by singing or by playing a musical instrument. The prototype Harmonizer is available on GitHub (Fernandes, 2022) and a video demonstrating its distinctive harmonizations is on YouTube (Link) as explained in the Results section of the paper.

## Contents

**1. Introduction**

In music, a melody consists of a series of notes played successively. Harmony is when two or more notes are played simultaneously. More specifically, harmony is the system of chords (simultaneous notes) and rules for the inter-chordal relationships that characterize Western music (Rich, 2022).

Melodic harmonization automatically assigns chords to a melody. It is difficult for a computer because it must understand harmony. I was motivated to prototype my own method of melodic harmonization because much of pop music nowadays sounds similar due to banal harmonies. Record companies have found a harmonic formula and discourage experimentation, with most popular songs even using the same four chords (Beato, 2017). I sought to create a fully-functioning, end-to-end system that inspires songwriters to create fresh harmonies.

In this paper, I explain "Harmonizer", my prototype system for melodic harmonization, which is available on GitHub (Fernandes, 2022). The completed application will use signal-processing to allow songwriters to record their own melody and provide them with a variety of unique auto-generated harmonizations.

Harmonizer takes a different approach from related methods in Section 2. Unlike those methods, Harmonizer uses Schoenberg's chart of regions (1969) as the underlying data structure that describes inter-chordal relationships. Since the chart reveals these relationships explicitly, Harmonizer explores harmonization methods used by practicing musicians. For example, applying the Church modes (Schroeder, 1998), melodizing awkward transitions with secondary dominants (Horton, 2020), and ii-V-I turnarounds (Workman, 2003).

**2. Related Work**

In their overview paper, Makris et al. (2016) explain two types of melodic harmonization: chorale harmonization and accompaniment. Chorale harmonization creates four-part harmonies for classical music. Accompaniment provides chord sequences to harmonize a melody. I focus on accompaniment because Harmonizer is intended for songwriters. Most accompaniment methods use a probabilistic framework to model harmony. I explain one such approach and then contrast it with the Harmonizer.

Songsmith (Simon, 2022), (Morris, 2008), and Band-in-a-Box (PG Music, 2022) are two commercial products that create accompaniments. The latter uses proprietary technology and studies show that it does not perform as well as the former (Simon, 2008). Songsmith auto-generates chords to accompany a vocal melody. It trains a Hidden-Markov-Model using a music database and applies that model to select chords. In Songsmith, the user must record a melody by singing along to a computer-generated beat at a user-specified tempo. Then, the system selects the best chord sequence, over 62 available chords, using the Viterbi algorithm (Forney, 1973).

Songsmith does melodic harmonization well (Microsoft, 2009). However, to produce a harmonious accompaniment, the Viterbi algorithm selects chords with the goal of minimizing the dissonance across all chord changes. While this does produce harmonious music, it is not what most practicing musicians do. They rely on musical theory to guide the chord selection. Although some chord selections are very dissonant and would be rejected by the Viterbi algorithm, skilled musicians can apply transitional chords to reduce that dissonance, producing fresher harmonies. Songsmith succeeds at creating a harmonization that likely fits the melody, but songwriters do not always want harmonizations that stick to safe chords; sometimes unexpected chord transitions add unique tones that complement the melody. This critique also applies to other accompaniment methods that use probabilistic frameworks to model harmony (Koops, 2012), (Kaliakatsos-Papakostas, 2014), (Weist, 2018).

Furthermore, Harmonizer also explores new signal-processing methods to enable songwriters to input melodies using musical instruments. This is more natural than the vocal input methods used by systems such as Songsmith. Section 3 describes the prototype Harmonizer system and Section 4 explains the various harmonization methods. Note that Appendix A explains some fundamentals of music theory and Appendix B contains some example files that are referenced throughout.

**3. System Description**

Figure 1 shows the three key components of the Harmonizer system: melody input, harmonization and playback. These three components are described in detail in the following sections. Harmonizer is primarily written in C++ and is available on Github (Fernandes, 2022). The melody-input and playback components use Javascript and Python code that is called from the C++ code. The prototype has only been tested on Mac OS. Section 6 discusses alternate execution modes.

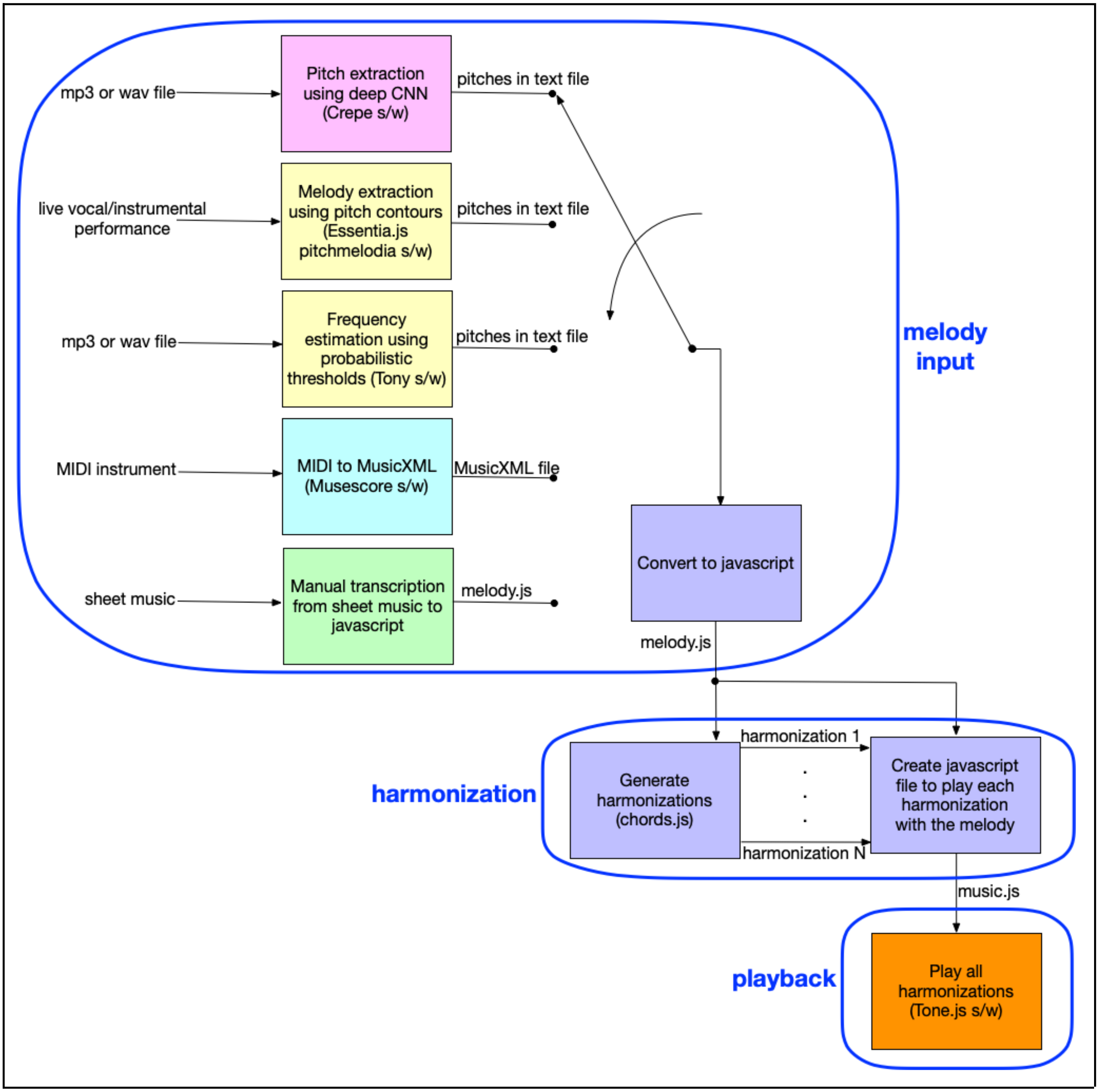

Figure 1: Block diagram of the prototype Harmonizer system

### 3.1 Melody input

The melodic input component consists of five methods to provide a melody. These five methods can be grouped into three categories: signal-processing methods, MIDI method and direct-transcription method. The signal-processing methods require little effort from the user. However, they sometimes detect erroneous pitches when the signal-to-noise ratio is low. The signal-processing methods are still a work in progress. Section 6 discusses how the signal-processing methods can be made more robust. To avoid dealing with false pitches, for development purposes, I added the MIDI and direct-transcription methods. These require more user-effort but provide perfect melodies.

#### 3.1.1 Signal-processing methods

##### 3.1.1.1 Melody extraction using pitch contours

I first attempted to extract the melody by singing or playing it on a musical instrument and recording it with the built-in microphone on my MacBook laptop. In Harmonizer, the main() function initiates this recording and saves it to an mp3 file. It calls a Javascript file, getPitches.js, to detect the pitches from the mp3 file via a pitch-detection algorithm. The detected pitches are saved to a file called pitches.txt. Appendix B contains an example fragment of pitches.txt. The first five entries in pitches.txt specify duration in seconds,  key of the song, major or minor mode, beats per minute and number of pitches. The remaining entries in the file specify detected pitches in Hertz. The pitches are equally spaced in time. After getPitches.js obtains pitches.txt, the prototype Harmonizer uses functions inputPitches() and formatPitches() to process pitches.txt and generate a Javascript file called melody.js which encodes the detected melody. An example fragment of melody.js is shown in Appendix B.

To obtain pitches.txt, I first tried the pitch-detection algorithm, PitchMelodia, from the Essentia.js open-source software package (Correya, 2021). PitchMelodia outputs pitches and associated confidence levels. It accurately detected pitches when the recording quality was clear and the melody was prominent over background noise. However, it gave very low confidence for pitch detection when the recording was not as clear, for example, if my laptop fan turned on.

I tried to improve the performance by attempting to understand how PitchMelodia works. From Salamon and Gomez's (2012) paper, PitchMelodia faces three challenges: determining when the melody is present (voicing detection), ensuring the estimated pitches are in the correct octave, and selecting the correct melody pitch when more than one note sounds simultaneously. The paper recommends applying a loudness filter to enhance frequencies that humans are sensitive to. It then converts the audio into the frequency domain via a Short-Term Fourier Transform. Further processing is on segments (hops) of the signal. I did not understand the math, but I think the algorithm tries to create a "pitch contour" which tracks the melody. Actually, there are several candidate contours and the most promising one is selected. Finally, the pitches are retrieved from the most promising contour.

From my limited understanding of PitchMelodia, it seems that the Essentia.js implementation does not apply a loudness filter. So, I tried to apply such a filter. I used the "loudness normalization" effect in the Audacity software package (Audacity, 2022). Unfortunately, that did not improve the performance. I also tried Salamon's Python code to convert the audio recording to MIDI but could not completely eliminate spurious pitches. Finally, I installed the PitchMelodia plugin into the Sonic Visualizer software package (Cannam, 2010). While this plugin did help me understand how PitchMelodia operates in the frequency domain, I was unable to further improve the performance.

#### 3.1.1.2 Melody extraction using deep convolutional neural networks

My second attempt at a pitch-detection algorithm was Kim et al.'s (2018) Crepe software package. Crepe uses a deep Convolutional Neural Network (CNN) to detect pitches from an audio signal. I did not fully understand the math in this machine-learning approach. The CNN weights are already trained from datasets in which the pitches are known (ground truth). Then, the trained CNN is used to detect pitches from the user's audio signal.

The Python code performed better than PitchMelodia. There were fewer spurious notes than in the PitchMelodia implementation. In Section 6, I describe some ideas to try in this approach.

#### 3.1.1.3 Melody extraction using probabilistic threshold distributions

Since my previous attempts at using signal-processing methods worked well but not perfectly, I decided to try an interactive method in which the user can correct the automatically detected pitches. The Tony (sic) software package (Tony, 2022), (March, 2014), provides a graphical interface that allows the user to do just that. If all the ideas in Section 6 fail to get nearly perfect results from the preceding signal-processing approaches, then I will use the Tony s/w in future Harmonizer versions, because interactive systems require slightly more user effort than non-interactive systems, but still less effort than non-signal-processing approaches.

### 3.1.2 MIDI method

To input a melody with a MIDI instrument, I used the MIDI-rec app (Kira, 2022) on an iPad. I played the melody on the MIDI keyboard in the app and then exported the MIDI files to the Musescore app (2022) for conversion to MusicXML format. A fragment of a file in MusicXML format (W3C, 2021) is shown in Figure 2. The prototype Harmonizer accepts the MusicXML as input and converts it to Javascript using the function getMelodyFromMusicxml() which implements the state-transition diagram in Figure 3.

To understand the state-transition diagram, observe that the MusicXML format in Figure 2 corresponds to the ordering of symbols in sheet music. I created the state-transition diagram as a representation of this ordering. For example, the divisions, fifths, beat-unit and per-minute tags must be received first to establish the key signature and timing information. Then the pitch, rest and other tags follow to specify notes and rests. In each state, internal information is updated and sometimes output to the melody.js file which encodes the melody in Javascript. An example fragment of melody.js is shown in Appendix B.

```
<?xml version="1.0" encoding="UTF-8"?>
…
<divisions>96</divisions>
<key>
      <fifths>-1</fifths>
</key>
<beat-unit>quarter</beat-unit>
<per-minute>120</per-minute>
<rest/>
  <duration>96</duration>
<rest/>
  <duration>12</duration>
<rest/>
  <duration>6</duration>
 <pitch>
      <step>C</step>
      <octave>4</octave>
  <duration>6</duration>
<tie type="start"/>
<pitch>
      <step>C</step>
      <octave>4</octave>
      <duration>12</duration>
<tie type="stop"/>
<rest/>
  <duration>12</duration>
...
```

Figure 2: foobar.musicXML

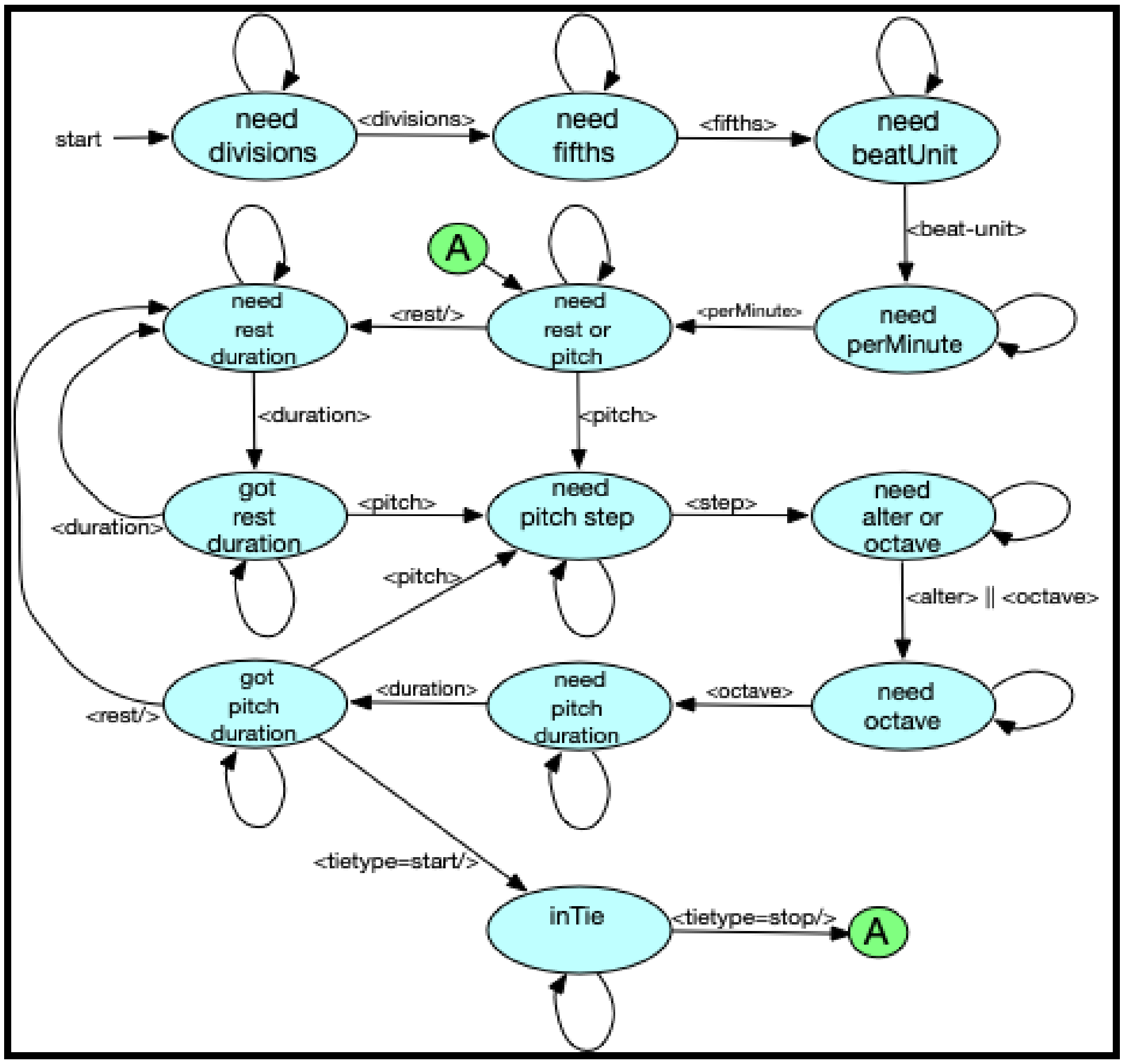

Figure 3: State-transition diagram to process musicXML and generate melody.js

### 3.1.3 Direct transcription

The last method for melody input serves more as a developer tool than melodic input. The user must directly add the information for their melody to an array in a javascript file. Each note in the melody must contain the following parameters: the relative time the note is played, the duration of the note, the note name (pitch class and octave), and the volume. Although this method provides a perfect method for melodic input, it is far too inconvenient for users.

### 3.2 Generating chord progressions for the melody

The harmonization methods are explained in Section 4. Each harmonization method is a function that returns a C++ vector of structs, in which each struct represents a chord. After all harmonization methods are called, the returned vectors are eventually added to a vector called “harmonizations”. Then, a function called

writeToChords() writes the chord structs in “harmonizations” into a Javascript file called chords.js. An example fragment of chords.js is shown in Appendix B. Finally, chords.js and melody.js are combined together into music.js, a Javascript file that contains all desired harmonizations along with the input melody. Appendix B shows an example fragment of music.js.

**3.3 Playback of the harmonizations**

To play the chords and melody from harmonizer.cpp, I use the Tone.js software package (Mann, 2015). The prototype Harmonizer invokes a browser to execute test.html, shown in Figure 4. Thus Tone.js plays the various harmonizations along with the melody that are both encoded in music.js. Note that music.js loads sampled mp3’s of all piano notes and then uses the Sampler functionality of Tone.js to play the harmonized melodies on a sampled grand piano.

```
<html lang="en">
<script src="../node_modules/tone/build/Tone.js"></script>
<script src="music.js"></script>
</html>
```

Figure 4: test.html

## 4. Harmonization Methods

In the following subsections, I describe the methods that I used to create "harmonization 1", …, "harmonization N" in Figure 1. The descriptions assume knowledge of music-theory fundamentals, which I have provided in Appendix A.

### 4.1 Simple harmonizations

For the first harmonization method that I created, I chose a simple yet effective harmonization principle: select the chord such that the melody note that it is harmonizing is the 3rd interval of the chord triad. Since the 3rd interval of a chord is the most emotive note in the chord, this chord always matches the melody. In this section, let us call such chords "matching chords". The note matches the chord because the 3rd note of a triad decides the chord's major/minor tonality. The other notes of the triad are simply taken from the scale of the melody. For example, to play a chord accompanying the note C in a melody in the key of  F Major, the chord {A, C, E} would be chosen. In this example, C is the third of the A minor (Am) triad.

The first harmonization, "Simple Harmonization 1" assigns main tones to notes. For this harmonization, a note is a main tone if it lasts longer than its 2 neighboring notes. Additionally, all tonic notes are main tones. Accidentals– notes not found in the melodic scale– can not be main tones. For every main tone, we simply play the matching chord of the note. The only exception is the tonic note, for which we always play the I chord of the scale.

The other simple harmonization, "Simple Harmonization 2", does not assign main tones. It plays a matching chord for every non-accidental note. The tonic is not treated differently. As a result, we play the vi/VI chord with the tonic note. This causes major melodies to sound a bit minor and minor melodies to sound a bit major.

**4.2 Harmonization using Schoenberg's regions**

When musicians analyze a musical composition, they may declare that it changes key (tonality) at certain points. However, in an attempt to establish how a composition can move through different "regions" within the SAME tonality, Schoenberg (1969) created the Chart of Regions in Figure 5. He insisted that moving along the chart of regions does not necessarily require modulation to other tonalities: "there is only *one tonality* in a piece, and every segment formerly considered as another tonality is only a region, a harmonic contrast within that tonality".

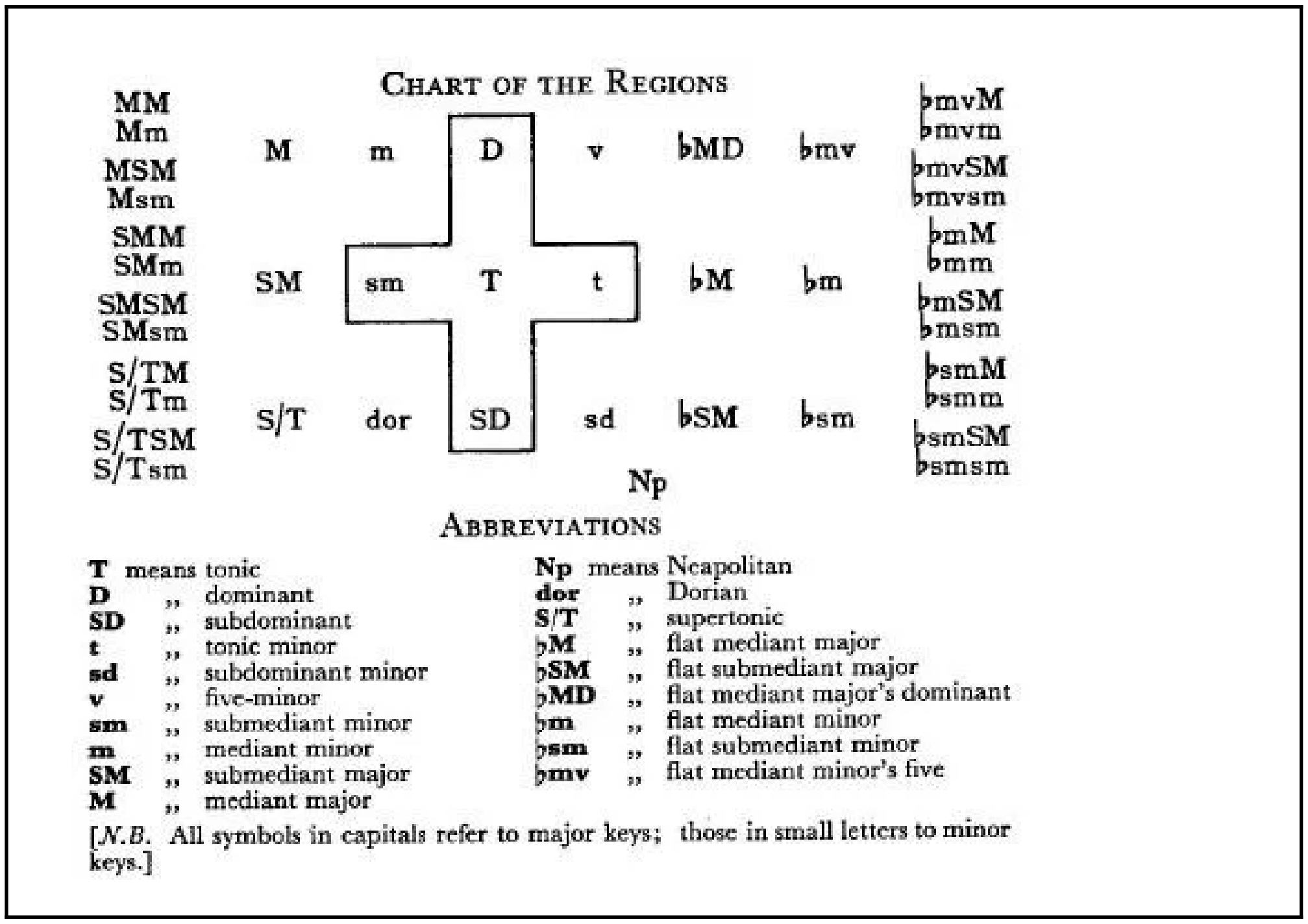

Figure 5: The chart of regions from Schoenberg's *Structural functions of harmony*.

Consider the three chords labeled T, D, SD, which is the region of the major chords in a chord scale (e.g., C, G, F, in C major scale). To the left of these are sm, m, dor, which is the region of the relative minors (e.g., Am, Em, Dm) and to the right of T, D, SD are v, t, sd, which is the region of the parallel minors (e.g., Cm, Gm, Fm). Thus the chart reveals key inter-chordal relationships given a tonic chord (e.g., C).

Although Schoenberg probably did not intend for the chart to be used beyond his region analysis, it does serve as a great tool for melodic harmonization. To do this, I expanded this chart and used it as the basis for many of my harmonizations. I created a 24x24 2D array with the melodic key at the center, shown in Figure 6, and then applied the inter-chordal relationships to populate all the chords.

| | | | | | | | | | | | | | | | | |
|---|---|---|---|---|---|---|---|---|---|---|---|---|---|---|---|---|
| B | b | D | d | F | f | G# | g# | B | b | D | d | F | f | G# | g# | B |
| E | e | G | g | A# | a# | C# | c# | E | e | G | g | A# | a# | C# | c# | E |
| A | a | C | c | D# | d# | F# | f# | A | a | C | c | D# | d# | F# | f# | A |
| D | d | F | f | G# | g# | B | b | D | d | F | f | G# | g# | B | b | D |
| G | g | A# | a# | C# | c# | E | e | G | g | A# | a# | C# | c# | E | e | G |
| C | c | D# | d# | F# | f# | A | a | C | c | D# | d# | F# | f# | A | a | C |
| F | f | G# | g# | B | b | D | d | F | f | G# | g# | B | b | D | d | F |
| A# | a# | C# | c# | E | e | G | g | A# | a# | C# | c# | E | e | G | g | A# |
| D# | d# | F# | f# | A | a | C | c | D# | d# | F# | f# | A | a | C | c | D# |
| G# | g# | B | b | D | d | F | f | G# | g# | B | b | D | d | F | f | G# |
| C# | c# | E | e | G | g | A# | a# | C# | c# | E | e | G | g | A# | a# | C# |
| F# | f# | A | a | C | c | D# | d# | F# | f# | A | a | C | c | D# | d# | F# |
| B | b | D | d | F | f | G# | g# | B | b | D | d | F | f | G# | g# | B |
| E | e | G | g | A# | a# | C# | c# | E | e | G | g | A# | a# | C# | c# | E |
| A | a | C | c | D# | d# | F# | f# | A | a | C | c | D# | d# | F# | f# | A |
| D | d | F | f | G# | g# | B | b | D | d | F | f | G# | g# | B | b | D |
| G | g | A# | a# | C# | c# | E | e | G | g | A# | a# | C# | c# | E | e | G |
| C | c | D# | d# | F# | f# | A | a | C | c | D# | d# | F# | f# | A | a | C |
| F | f | G# | g# | B | b | D | d | F | f | G# | g# | B | b | D | d | F |
| A# | a# | C# | c# | E | e | G | g | A# | a# | C# | c# | E | e | G | g | A# |
| D# | d# | F# | f# | A | a | C | c | D# | d# | F# | f# | A | a | C | c | D# |
| G# | g# | B | b | D | d | F | f | G# | g# | B | b | D | d | F | f | G# |
| C# | c# | E | e | G | g | A# | a# | C# | c# | E | e | G | g | A# | a# | C# |
| F# | f# | A | a | C | c | D# | d# | F# | f# | A | a | C | c | D# | d# | F# |
| B | b | D | d | F | f | G# | g# | B | b | D | d | F | f | G# | g# | B |

Figure 6: A section of the 2D array chart of regions. In each column, the chords cycle through the circle of fifths. In each row, the chords move from relative minor on the left to major to parallel minor on the right. Minor chords are lowercase, major chords are uppercase.

On Pg. 22, Schoenberg (1969) states that “modulation from one region to another … is based on at least one chord common to both regions”. Therefore, we can harmonize melodies by traversing either horizontally or vertically across the chart in single steps. Such traversals move across the regions (circle of fifths, major region, relative-minor region, parallel-minor region) with harmonious chord changes. Other traversals (e.g., diagonal or large steps) can sound less harmonious. However, the final paragraph of Appendix A explains how secondary dominants and ii-V-I turnarounds can make such awkward-sounding transitions sound mellow.

In the prototype Harmonizer, the schoenberg() function creates harmonizations by traversing through the chart. These harmonizations always start and end on the initial (tonic) chord of the melody. The function accepts many parameters including the possible directions to travel, whether repeats are allowed, whether the harmonization can change mode, the probability of using a secondary dominant, and the probability of using a ii-V-I. As a result, many interesting methods of harmonization can be made from this one function. I have created 2 simple ones, showing the diversity of options available: SchoenbergMin and SchoenbergMax.

SchoenbergMin allows traversal from the current key to its 4 immediate neighbors. Additionally, this harmonization always uses secondary dominants and has a 25% chance of using a ii-V-I turnaround. Each chord is repeated 4 times per measure.

SchoenbergMax can traverse 12 directions instead of 4. These directions are the 4 immediate neighbors; the 4 diagonals; and 2 steps up, 2 steps down, 2 steps left, and 2 steps right. Chords are only played once, and secondary dominants are not always used, causing this harmonization to sound more varied than the previous.

**4.3 Harmonization using Giant Steps**

John Coltrane's classic jazz piece "Giant Steps" uses a creative approach to harmonization where the song modulates between 3 keys, each 4 steps from the other on the circle of fifths. My GiantSteps harmonization replicates this progression, using the schoenberg() function to only move 4 steps up from the tonic, emulating a path across the circle of fifths. This harmonization always uses secondary dominants and has a 25% chance of a ii-V-I turnaround. Every chord is only repeated once to create a rapid, chaotic feel.

**4.4 Harmonization using church modes**

As explained in Appendix A, excluding the ionian major and aeolian minor, there are 5 church modes: dorian, phyrgian, lydian, mixolydian, and locrian. Lydian and mixolydian are major modes while the other 3 are minor. Since every mode has a different special note, I did not create a definitive method of chord progression for each mode. Instead, each mode simply has its own unique chord progression with the goal of highlighting the modal center and the mode's special note (Beatham, 2022).

The first mode, dorian mode, is a minor mode and has a sharp 6th as its special note. Since dorian is so close to ionian, it is common to think of the dorian mode as ionian, but with an emphasis on the ii chord. The dorian chord progression I chose is ii, iii, ii, and V. The V chord exists to create tension and also highlights the dorian mode's sharp 6th (the 7th note of the ionian). The iii chord spices up the progression.

The second mode, phrygian mode, proved rather difficult to harmonize, so I used phrygian dominant instead. Phrygian has both a minor second and minor third, which often conflict. Phrygian dominant, on the other hand, has a major third. This means that both the I and II chord, only one half-step away from each other, are major, giving it its unique sound. I used the progression I, II to achieve a very mysterious sound common in middle eastern and flamenco music.

The third mode, lydian mode, is a common major mode with a sharp fourth as its special note. To highlight this special note, the lydian harmonization includes the II chord, which becomes major due to the sharp fourth. I use the progression I, II for lydian.

The fourth mode, mixolydian mode, is a major mode with a flat 7th. To highlight this special note, we play VII and I.

The fifth mode, locrian mode, proves extremely dissonant on its own. As such, it is commonly used to transition to ionian or aeolian mode. Similar to the way I treated dorian mode, I just think of locrian mode as ionian with a focus on the vii chord. The progression simply transitions between vii and I.

**5. Results and Discussion**

I tested the Harmonizer prototype with two well-known melodies: Happy Birthday (HB) and Für Elise (FE). HB is in the key of F major and FE is in A minor. I played both melodies on the MIDI-rec app (Kira, 2022) on an iPad and then exported MIDI files to Musescore (2022) for conversion to MusicXML format. Harmonizer then converted the MusicXML format to javascript as explained in Section 3.1.2 and harmonized the HB and FE melodies using the methods of Section 4. I created a video of the melodic input process and playback of the harmonizations and uploaded it to YouTube (Link). Timestamps are provided below and in the YouTube video description.

For HB, the harmonizations use the chord progressions and appear at the timestamps shown below:

1. Simple harmonization 1 (2:24)

`C C F C C C F C F Gm F F`

2. Simple harmonization 2 (3:39)

`Am Am Bb Am Dm C Am Am Bb Am Edim Dm Am Am Am F F Dm C Bb C Bb Gm Gm F Dm Edim Dm`

3. SchoenbergMin (1:09)

F F F F Cm7 Fm Fm Fm Fm Fm7 Bbm Bbm Bbm Bbm Ab7 Db Db Db Db Eb7 A A A A C7

F F F

4. SchoenbergMax (3:14)

F Am7 Dm Cm C C7 F D7 G Dm E7 A C#7 F# Eb Ebm7 Abm Bbm7 Ebm Fm7 Bbm Bb C

Bbm7 Eb Cm7 Fm Cm Am A7 D C C7 F

5. GiantSteps (1:58)

F Ebm Ab7 Db Db E7 A Gm C7 F Ebm Ab7 Db Db E7 A Gm C7 F Ebm Ab7 Db Db E7

A A C7 F F C7 F

6. Lydian harmonization (1:33)

F F G7 G7 F F G7 G7 F F F F

7. Mixolydian harmonization (2:49)

F7 F7 Eb Eb F7 F7 Eb Eb F7 F7 F7 F7

For FE, the harmonizations appear at the following timestamps in the video:

1. Simple harmonization 1 (5:56)

Em Em Em Am Em Am G Em Am Em Em Em Am Em Am G Em Am

2. Simple harmonization 2 (7:45)

C C C G Bdim Am F Am C F G C G Am C C C C G Bdim Am F Am C F G C Am G F

3. SchoenbergMin (4:35)

Am Am Am Am G7 C C C C C7 F F F F Cm7 Fm Fm Fm Fm Gm7 Cm Cm Cm Cm E7 Am Am

Am

4. SchoenbergMax (6:50)

Am E7 A Em Em7 Am F#m7 B A Abm7 C#m Fm7 Bbm Gm Bb7 Eb F7 Bb C7 F Dm G Dm7

Gm G7 C G E C#m7 F#m Em Em7 Am

5. Giant Steps (5:28)

`Am Gdim Cm7 Fm Fm Ab7 C#m Bdim Em7 Am Gdim Cm7 Fm Fm  Abm7 Dbm Bdim Em7 Am Gdim Cm7 Fm Fm Abm7 C#m C#m Em7 Am Am Em7 Am`

6. Dorian harmonization (6:23)

`Am B Am D Am B Am D Am B Am`

7. Phrygian dominant harmonization (5:01)

`Bb Bb A A Bb Bb A A Bb Bb Bb Bb`

8. Locrian harmonization (7:18)

`Bb Bb Adim Adim Bb Bb Adim Adim Bb Bb Bb Bb`

For both melodies, the simple harmonizations sound conventional and similar to Songsmith harmonizations. The modal harmonies (Lydian, Mixolydian, Dorian, Locrian and Phrygian dominant) are also conventional but different from Songsmith harmonizations. These are intended to inspire songwriters to consider modal harmonies. All modal harmonies do not work equally well over a given melody. If a melody uses most notes from a particular modal scale, and few notes outside that scale, then that modal harmonization will sound good over the melody. For example, over the HB melody, the Mixolydian harmonization sounds better than the Lydian harmonization and the Phrygian dominant harmonization works best over the FE melody.

For both HB and FE melodies, the Giant Steps, SchoenbergMin and SchoenbergMax harmonizations are unconventional and completely different from Songsmith harmonizations. These harmonizations progress through a large number of chord changes, but the use of secondary dominants and ii-V-I turnarounds ensure that they never sound discordant. In particular, the Giant Steps and SchoenbergMax harmonizations move through several chords that are musically unexpected in the keys of F major and A minor. These harmonizations show that the Schoenberg regions can indeed be used as a data structure for melodic harmonization and that it allows the application of the concepts that practicing musicians use.

**6. Future Work**

Overall, the prototype Harmonizer serves as a foundation for a multitude of possibilities. The following subsections describe improvements that I intend to pursue. I invite interested readers to provide feedback and to experiment and contribute to the prototype Harmonizer on GitHub (Fernandes, 2022). Although Harmonizer could be a great commercial product, I want it to be accessible to any songwriter. This can be done by releasing the final version on computer and mobile platforms.

**6.1 Signal-processing improvements**

As explained in Section 3.1.1, the signal-processing approach to inputting a melody will greatly enhance the user-experience. In that section, I explained my initial experiments with PitchMelodia and Crepe. These methods work well, but I could not get them to work perfectly. I plan to try the following experiments:

1. Use an external microphone instead of my built-in MacBook laptop microphone to reduce noise which may interfere with the pitch detection.
2. Explore noise-reduction and loudness normalization filters (e.g., replayGain).
3. Try changing the PitchMelodia parameters.
4. Post-process the pitches.txt file to filter out spurious frequencies with a median filter that removes outliers.
5. Eliminate pitches with low confidence values and replace them with adjacent pitches with higher confidence values.
6. In all preceding experiments, examine the frequency-domain waveforms in Audacity and use that information to guide the experiments.

### 6.2 Harmonization improvements

The harmonization approaches in Section 4 illustrate the potential of Schoenberg's chart of regions. I plan to try the following experiments:

1. Apply Schoenberg's analysis of regions to create more unconventional, but grounded harmonizations. In Chapter IX, Schoenberg (1969) classifies the relationships between regions as follows:
    a. Type a: direct and close
    b. Type b: indirect but close
    c. Type c: indirect and remote
    d. Type d: distant

It would be interesting to experiment with a traversal of the regions in which the most frequent transitions are of Type a, then Type b and Type c. Type d transitions should occur rarely. Combined with secondary dominants and ii-V-I turnarounds this harmonization should sound conventional but it will also contain many interesting chord changes. Varying the occurrence frequencies of the different types will influence the perceived effect of the music.

2. Instead of a random traversal of the chart, the melody notes should guide the traversal. Schoenberg provides numerous examples of how the great composers apply transitions of Types a, b, c, d, based on the underlying melody notes.
3. Create multiple progressions for each modal harmonization. Ideally, the notes of the melody should be analyzed to determine which modal chords are the best fit.
4. Improve the voicing of the harmonizations to better match the melody. This can be done by creating a function that automatically rearranges notes in a series of chords to match a melody.

### 6.3 User-interface improvements

I plan to implement these user-interface improvements:

1. Printing the chords in sheet music along with their names.
2. Creating stand-alone computer and mobile apps.

**Bibliography**

Audacity Team. "Audacity(R): Free Audio Editor and Recorder [Computer application]." Version 3.1.3 retrieved June 27th 2022, www.audacityteam.org.

Beatham, Mike. “Learn guitar scales.” *Fretjam*, Accessed Jul. 5, 2022, www.fretjam.com/guitar-scales.html.

Audacity Team. "Audacity(R): Free Audio Editor and Recorder [Computer application]." Version 3.1.3 retrieved June 27th 2021, www.audacityteam.org.

Beato, Rick. “The four chords that killed POP music.” *YouTube*, Uploaded by Everything Music, Dec. 14, 2017, www.youtube.com/watch?v=nuGt-ZG39cU&list=PLW0NGgv1qnfyK8S0RWhsFT8-U0di93oRo&index=5.

Cannam, Chris, Christian Landone and Mark Sandler. "Sonic Visualiser: An Open Source Application for Viewing, Analyzing, and Annotating Music Audio Files." *Proceedings of the ACM Multimedia 2010 International Conference*, pp. 1467-68, October 2010, Firenze, Italy.

Correya, Albin, Jorge Marcos-Fernández, Luis Joglar-Ongay, Pablo Alonso-Jiménez, Xavier Serra, Dmitry Bogdanov. “Audio and Music Analysis on the Web using Essentia.js.” *Transactions of the International Society for Music Information Retrieval (TISMIR)*, 4(1), pp. 167–181. 2021.

Fernandes, Tony. “Harmonizer Repository on Github.” github.com/desanton/Harmonizer, Aug. 17, 2022

Forney, G. David. "The Viterbi Algorithm." Proceedings of the IEEE 61.3 (1973): 268-278.

Horton, Charles, David A. Byrne, and Lawrence Ritchey. “Harmony through melody: The interaction of melody, counterpoint, and harmony in Western music.” Rowman & Littlefield Publishers, 2020.

Kaliakatsos-Papakostas, Maximos, and Emilios Cambouropoulos. "Probabilistic harmonization with fixed intermediate chord constraints." *ICMC*. 2014.

Kim, Jong Wook, Justin Salamon, Peter Li, and Juan Pablo Bello. "Crepe: A convolutional representation for pitch estimation." In *2018 IEEE International Conference on Acoustics, Speech and Signal Processing (ICASSP)*, pp. 161-165. IEEE, 2018.

Kira, Ryouta. "MIDI Recorder with E.Piano", Apple App Store, Accessed Jul. 25, 2022, apps.apple.com/us/app/midi-recorder-with-e-piano/id1448577506

Koops, H. V. "A model based approach to automatic harmonization of a melody." Bachelor's thesis, Utrecht University, 2012.

Mann, Yotam. "Interactive music with tone. js." *Proceedings of the 1st annual Web Audio Conference*. Citeseer, 2015.

Makris, Dimos, Ioannis Kayrdis, and Spyros Sioutas. "Automatic melodic harmonization: An overview, challenges and future directions." *Trends in music information seeking, behavior, and retrieval for creativity* (2016): 146-165.

Mauch, Matthias, and Simon Dixon. "pYIN: A fundamental frequency estimator using probabilistic threshold distributions." In 2014 IEEE International Conference on Acoustics, Speech and Signal Processing (ICASSP), pp. 659-663. IEEE, 2014.

Microsoft Research. “Microsoft Songsmith Commercial.” YouTube, Uploaded by Duker91, Jan. 8, 2009, www.youtube.com/watch?v=3oGFogwcx-E

Morris, Dan, Ian Simon, and Sumit Basu. "Exposing parameters of a trained dynamic model for interactive music creation." In *AAAI'08 Proceedings of the 23rd national conference on Artificial intelligence-Volume 2*, pp. 784-791. 2008.

Musescore. “The world's most popular notation software.” Accessed Jul. 15, 2022, www.musescore.org

PG Music Inc: Band-in-a-Box. www.pgmusic.com. Accessed Aug., 10, 2022

Rich, Alan. "Harmony". Encyclopedia Britannica, www.britannica.com/art/harmony-music. Accessed 13 Aug. 13, 2022.

Salamon, Justin, and Emilia Gómez. "Melody extraction from polyphonic music signals using pitch contour characteristics." *IEEE transactions on audio, speech, and language processing* 20.6 (2012): 1759-1770.

Schoenberg, Arnold, and Leonard Stein. *Structural functions of harmony*. No. 478. WW Norton & Company, 1969.

Schroeder, Carl, and Keith Wyatt. Harmony and Theory: A Comprehensive Source for All Musicians. Hal Leonard Corporation, 1998.

Simon, Ian, Dan Morris, and Sumit Basu. "MySong: automatic accompaniment generation for vocal melodies." In *Proceedings of the SIGCHI conference on human factors in computing systems*, pp. 725-734. 2008.

Simon, Ian and Sumit Basu. “Songsmith: Automatic accompaniment for vocal melodies.” Microsoft Research, Accessed Jul. 5, 2022, www.microsoft.com/en-us/research/project/songsmith-2/

Tony team. "Tony: a tool for melody transcription." Retrieved Jul. 10, 2022. code.soundsoftware.ac.uk/projects/tony

W3C Community Group. "MusicXML 4.0: final community group report." Jun. 2021, www.w3.org/2021/06/musicxml40/

Weist, Fraser. "Order Out of Chaos: Randomized Heuristics and Their Application to the Harmonization Problem." Bachelor’s thesis, Harvard University, 2018.

Workman, Josh. “Chops: II–V–I Survival Tips.” *Guitar Player*, Apr. 2003, pp. 90–100.

**Appendix A: Music theory fundamentals**

In classical western music theory, we have twelve pitch classes: {C, C#/Db, D, D#/Eb, E, F, F#/Gb, G, G#/Ab, A, A#/Bb, B}. Note that some sharp notes equal their flat counterpart, such as C# and Db. For simplicity, Harmonizer does not use flat notes.

The piano has several octaves, each containing the twelve pitch classes. To increase the octave and stay within the pitch class, we must double the frequency. Therefore, every note on the piano can be described by the pitch class and octave. For example, the standard pitch notation for the A pitch class in the 4th octave would be A4. The interval between adjacent pitch classes is called a half-step/semitone. The interval between 2 steps is a whole-step/tone.

In tonal music (most popular music), we designate a tonal center called the *tonic*. From the tonic, we conceptualize the relationship to other pitch classes using *scales*. Different scales have different interval patterns. The 2 most common scales are major and minor. The major scale generally produces a happy sound while the minor scale produces a sad sound. The major scale has an interval pattern of whole step, whole step, half step, whole step, whole step, whole step, half step. Therefore, if F is the tonic, then F major scale has pitch classes {F, G, A, A#/Bb, C, D, E, F}. Note that the scale ends on F and can therefore be repeated upwards. The minor scale has an interval pattern of whole step, half step, whole step, whole step, half step, whole step, whole step. The F minor scale is {F, G, G#/Ab, A#/Bb, C, C#/Db, D#/Eb, F}.

From the major scale we have 7 *church modes*: Ionian (major), dorian, phrygian, lydian, mixolydian, aeolian (minor), and locrian. These modes are built from the pitch classes of the major scale, but have a different tonic center. For example, the dorian mode is the 2nd mode meaning its tonic center is the 2nd pitch class of the relative major scale. F major would have G dorian: {G, A, A#/Bb, C, D, E, F, G}. These modes have unique interval patterns, these patterns being variations of the major interval pattern.

Since each mode of a scale is comprised of the same pitch classes, it is important to emphasize a mode's tonic center. In addition to the tonic center, each mode has a special alteration from the major scale that makes it sound unique. For example, the difference between C Major and C lydian is that lydian has a F#/Gb instead of an F. This sharp fourth is the special note of a lydian scale.

Chords are collections of notes played at the same time. Ultimately, my harmonizations are just a series of chords. Chords generally have a root note and the other notes of the chord are relative to the root note. Chords can be built from scales. When doing so, the intervals between notes in these chords are the number of steps between these notes in a scale. For example, if we build a chord from F Major, then G would be the 2nd, A the 3rd, and so on.

My harmonizations are built from 2 types of chords: triads and sevenths. A triad is a chord built from the root note, the 3rd from the root, and the 5th. A major triad has a 3rd that is 4 semitones above the root note and a 5th that is 7 semitones above the root note. A minor triad is like a major triad, but with a 3rd 3 semitones above the root note. A diminished triad is like a minor triad, but with a 5th 6 semitones above the root note. A seventh chord is a triad but with an extra note: the 7th. The 7th is either 10 semitones or 11 semitones above the root note. When it is 10 semitones, it is called a dominant 7th.

The root note is not necessarily the lowest note on the piano. A chord is truly made up of the pitch class relations between the notes, meaning the notes can be oriented in any way and we would call it the same chord. For example, an A major triad is made up of {A, C#, E}. If we play it as A4, C#6, E2, it would still be an A major triad, even if E is the lowest note. Changing the ordering of a chord is called an *inversion*. Depending on the melody and other chords, we may want different notes to be at the bottom or at the top.

To denote the chord of a scale, it is common to use the roman numeral. Additionally, if the chord has a major 3rd, then the numeral is capitalized. (For example, the first, fourth and fifth major chords are I, IV and V. In F major, these would be F, A#/Bb, C.) If the chord has a minor 3rd, the numeral is left lowercase. So, the second, third, sixth and seventh minor chords are ii, iii, vi, vii. In F major, these would be Gm, Am, Dm, Em (or E diminished).

Finally, I shall briefly describe the concepts of secondary dominants (Horton, 2020), and ii-V-I turnarounds (Workman, 2003). Using the preceding notation,  the seven chords of a chord scale are I, ii, iii, IV, V, vi, vii. In F major, these chords are F, Gm, Am, Bb, C, Dm, Edim. Now the V chord is C in the key of F major. The notes of the V chord are C-E-G and those of the I chord are F-A-C. If we build the chords with 4 notes, then they are C-E-G-Bb and F-A-C-E. In the C dominant 7th chord, the interval between E and Bb is a tritone and it gives the listener a suspenseful feeling because the relative frequencies have a ratio of 32/45 which sounds unstable compared to the stable ratios of 2/3 (C to G) and 4/5 (C to E). If the F-A-C or F-A-C-E chord immediately follows the C-E-G-Bb, then the suspenseful feeling is resolved because the tritone is absent and the consecutive chords have C and E notes in common and the E-Bb notes are one semitone away from the F-A notes. Thus the suspenseful feeling has been smoothly resolved. In practice, an awkward-sounding transition to an F chord may be made more harmonious by preceding the F chord with its secondary dominant, the C7. The listener will experience a momentary, suspenseful feeling but that will be smoothly resolved by the F chord. In general, an awkward-sounding transition to any chord may be made more harmonious by preceding that chord with its secondary dominant V chord to get a harmonious V-I transition. The ii-V-I turnaround is an extension of the V-I transition in which the V chord is preceded by a ii chord. Using 4-note chords in the key of F major, the ii-V-I turnaround is Gm7 to C7 to Fmaj7 which is G-Bb-D-F to C-E-G-Bb to F-A-C-E. The Gm7 to C7 transition sounds smooth because G-Bb are common notes and the F and E are a semitone away. This smooth transition is then immediately followed by a secondary dominant resolution. Thus the ii-V-I turnaround also makes awkward-sounding transitions more harmonious by using a secondary dominant preceded by a smooth transition.

## Appendix B: Example files for the Happy-Birthday melody

**Pitches.txt:**

```
84.02 F major 114.84 7892 394.344 393.742 393.74 393.289 392.612 …
```

**Melody.js:**

```
const melody = [
      {'time': '4.000000',   'note': 'C5', 'duration': '0.750000'},
      {'time': '4.750000',   'note': 'C5', 'duration': '0.250000'},
      {'time': '5.000000',   'note': 'D5', 'duration': '1.000000'},
      …
];
const key = 'F';
const mode = 'major';
```

**Chords.js:**

```
const chords = [
      {'time': '4.000000', 'note': ['A2', 'C3', 'E3', 'A3'], 'duration': '1.000000’},
      {'time': '5.000000', 'note': ['A2', 'C3', 'E3', 'A3'], 'duration': '1.000000'},
      {'time': '6.000000', 'note': ['A2', 'C3', 'E3', 'A3'], 'duration': '1.000000'},
      {'time': '7.000000', 'note': ['A2', 'C3', 'E3', 'A3'], 'duration': '0.500000'},
      …
];
```

**Music.js:**

```
// Tone.sampler section
```

```
const sampler = new Tone.Sampler({
   urls: {
       "A7": "A7.mp3",
       "A1": "A1.mp3",
       "A2": "A2.mp3",
       "A3": "A3.mp3",
…
   },
   baseUrl: "pianoSamples/"
}).toDestination();

Tone.Transport.bpm.value = 150

const melody = [
      {'time': '4.000000',   'note': 'C5', 'duration': '0.750000'},
      {'time': '4.750000',   'note': 'C5', 'duration': '0.250000'},
      {'time': '5.000000',   'note': 'D5', 'duration': '1.000000'},
      …
];
const key = 'F';
const mode = 'major';

const chords = [
      {'time': '4.000000', 'note': ['A2', 'C3', 'E3', 'A3'], 'duration': '1.000000'},
      {'time': '5.000000', 'note': ['A2', 'C3', 'E3', 'A3'], 'duration': '1.000000'},
      {'time': '6.000000', 'note': ['A2', 'C3', 'E3', 'A3'], 'duration': '1.000000'},
      {'time': '7.000000', 'note': ['A2', 'C3', 'E3', 'A3'], 'duration': '0.500000'},
      …
];

Tone.loaded().then(() => {
   const part = new Tone.Part(function(time, value){
       sampler.triggerAttackRelease(value.note, value.duration, time);
   }, chords).start(0);
})

const part = new Tone.Part(function(time, note){
```

```
   sampler.triggerAttackRelease(note.note, note.duration, time);
}, melody).start(0);

part.humanize = true;
Tone.Transport.start();
```